# Positive and negative charge trapping GaN HEMTs: interplay between thermal emission and transport-limited processes

A. Nardo[a,*], C. De Santi[a], C. Koller[b], C. Ostermaier[b], I. Daumiller[b],
G. Meneghesso[a], E. Zanoni[a,] M. Meneghini[a]

[a] *Department of Information Engineering, University of Padova, via Gradenigo 6/B, 35131 Padova, Italy*
[b] *Infineon Technologies Austria AG, Siemensstraße 2, 9500 Villach, Austria*

**Abstract**

   This paper investigates the kinetics of buffer trapping in GaN-based normally-off high-voltage transistors. The analysis was carried out on transfer-length method (TLM) structures.
By means of a custom setup, (i) we investigated the trapping and de-trapping processes induced by a large vertical bias and identified different mechanisms, responsible for the storage of negative and positive charge in the buffer. (ii) temperature-dependent analysis was carried out to evaluate the time constants associated to negative and positive charge build-up. Remarkably, the results indicate that the activation energy for negative charge trapping is ~0.3 eV, which is much lower than the ionization energy of carbon acceptors (0.8-0.9 eV). This result is explained by considering that trapping and de-trapping are not dominated by thermal processes (thermal emission from acceptors), but by transport mechanisms, that limit the transfer of charge to trap states. (iii) in the recovery experiments, after low stress bias negative charge trapping dominates. After high stress bias, also the effect of positive charge generation is detected, and the related activation energy is evaluated.
The results presented within this paper clearly indicate that the trapping and de-trapping kinetics of normally-off GaN HEMTs are the results of the interplay of transport-limited conduction processes, that result in a low thermal activation (Ea~0.3 eV), compared to that of $C_N$ acceptors.

## 1. Introduction

Due to its superior material properties compared to silicon (high maximum current, high breakdown voltage, and high switching frequency [1]–[3]), gallium nitride represents the material of choice to increase the efficiency of next-generation power converters. Systems employing GaN based devices have higher efficiency, corresponding to lower losses, and higher switching frequencies, that allow to reduce the size and weight of the converters. In order to develop robust and reliable GaN-based power devices, it is necessary to understand the physical mechanisms that influence the operation and the lifetime of devices and their link to structure and material properties. One key requirement to optimize GaN-based power transistors is an appropriate choice and design of the buffer. In order to compensate the unintentional n-type conductivity in GaN HEMTs, carbon is usually used as dopant for power switching applications [4], [5]. Although carbon doping results in excellent breakdown properties, it has also been linked with current-collapse effects. This represents a challenging trade-off [6]–[10], and may significantly impact on device performance [4], [11], [12].

The goal of this paper is to provide further insight on the origin of positive and negative charge storage in the buffer of GaN HEMTs: we demonstrate that the dynamic response of the devices depends on the interplay between transport-limited charge storage/release mechanisms, and provide general conclusions, that can be used for understanding and optimizing the dynamic behavior of the devices.

## 2. Experimental details

The epitaxial structure consists of a AlN/AlGaN/GaN strain relief layer grown on a highly p-doped silicon substrate, a C-doped GaN layer, an unintentionally-doped GaN channel layer and an AlGaN barrier layer. Carbon doping in the buffer is used to counter dope the intrinsic n-type conductivity of GaN buffer after MOCVD growth. High concentration ($10^{19}$ cm$^{-3}$) is used in the GaN buffer to this aim. Studying buffer trapping phenomena in state-of-the-art devices is not straightforward. We decided to carry out our analysis on transfer-length method (TLM) structures, that have the same epitaxial structure of the HEMTs. All trapping processes identified in this paper can then be ascribed to the buffer, and are investigated with much more accuracy than working on HEMTs.

---

* Corresponding author. nardoari@dei.unipd.it

## 3. Results and Discussion

*3.1 Time-dependent measurements*

In order to gain insight on the trapping and leakage transport mechanisms in the C-doped GaN buffer layer, we developed a custom measurement based on the use of a semiconductor parameter analyzer (Fig. 1). This measurement captures the time response of the GaN:C buffer, and allows by a relatively simple approach to establish the mechanisms that are dominant, looking at both the conductivity of the device and the leakage current.

The measurement is made up by two phases:

-FILLING PHASE: the conductivity of the 2-DEG is measured by using a small bias of <1 V between two ohmic contacts, while a constant voltage $V_B$ is applied to the Si-substrate. This corresponds to the polarity experienced by a transistor under OFF state conditions (vertical field between drain and substrate).

-RECOVERY PHASE: the constant filling voltage is removed and the recovery of the conductivity of the 2DEG is monitored for 1000s.

In this analysis we have applied to the device a $V_{DS}$=0.1V and different $V_B$, from -10V up to -200V.

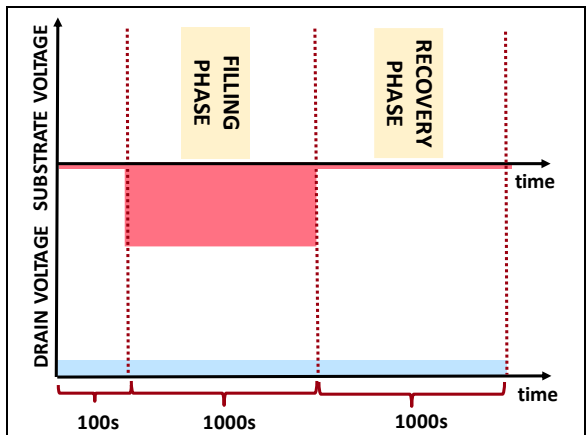

Fig. 1: Schematic representation of the experimental setup used for time-dependent measurements.

Fig. 2 shows the time-dependent evolution for two selected temperatures, 50C and 150C, corresponding to low and high temperature regime.

When the negative filling bias is applied, the drain current decreases due to the electrostatic effect of the backside potential. This effect is almost instantaneous, and may be followed by slower charge trapping/de-trapping phenomena, that typically have the shape of (stretched) exponentials. Any changes in the conductivity can be used to quantify bulk charge storage and trapping. At the lowest voltages (-10/-20 V in Fig. 2), a small negative going conductivity transient is observed during the first seconds of substrate filling, whereas at larger negative voltages (-100 V and below), this is overlaid by a positive going transient. The negative transient corresponds to negative charge storage in the buffer, whereas the positive transient corresponds to positive charge storage.

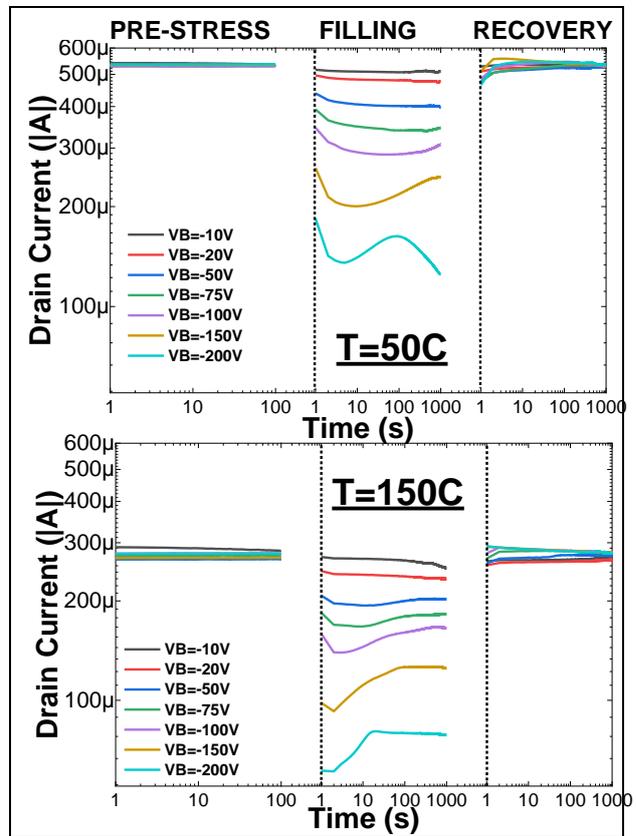

Fig. 2: Transient behavior of drain current in the low and high temperature regime.

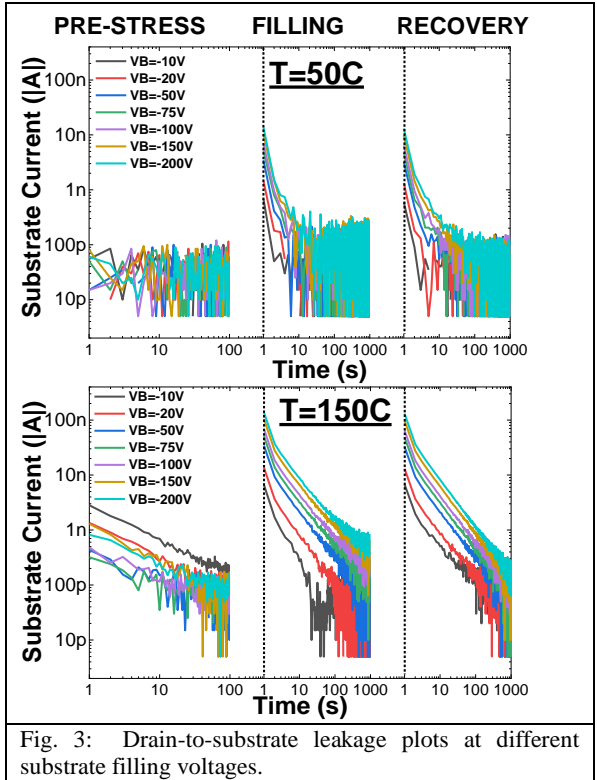

Fig. 3: Drain-to-substrate leakage plots at different substrate filling voltages.

Significant qualitative/quantitative differences are evident when comparing measurements in the low and high temperature regime. Temperature accelerates the two mechanisms, confirming we are dealing with charge-trapping processes and not with transients related to the charging of parasitic capacitances of the device. This is clearly evident in the drain-to-substrate leakage plots (Fig. 3. At high temperature, substrate-current transients are observed: a similar amount of carriers are injected and extracted from the structure, leading the device back to rest condition. This is a first indication that vertical carrier transport (injection/ejection of electrons from/to the substrate) can play a major role during a backgating experiment.

*3.2 Interpretation: different trapping mechanisms*
The presence of positive and negative current transients is explained by considering recent literature on this topic [7]–[9], [13]. The model is based on the fact that intentional buffer doping with carbon leads to $C_N$ acceptor traps located in the lower half of the bandgap. Even if, as reported in [14], the concentration of holes is very low and they are isolated from 2DEG by a depletion region, they can influence charge trapping processes and be stored as free carriers at the bottom of the stack. In our hypothesis, when the external electric field is applied to the GaN:C layer, three mechanisms can contribute to charge trapping, as shown in the schematic in Fig. 4:

-Process 1: *charge redistribution in C-doped GaN*. Due to an electric field in the C-doped GaN layer, positive charges will propagate from the top of C-doped GaN to the bottom. This will leave a negative charge (fixed acceptor) at the top of the C-GaN layer, and a positive charge at the bottom of the buffer. The positive charge coming from the top can either be (Process 1a) a hole that is emitted from a carbon acceptor to the valence band and propagates then in the valence band, or (Process 1b) a positive charge propagating within a defect band in the band gap through C-GaN states, see [14]. Hole emission (Process 1a) should show an activation energy of 0.8-0.9 eV, which might decrease at high electric fields due to Poole-Frenkel effect. On the other hand, for defect band conduction (Process 1b) Koller et al. [14] claimed that this process is not Arrhenius-like. As a consequence, the activation energy extracted from Arrhenius plots differs from 0.8-0.9 eV and is unphysical (i.e. not related to a carrier emission process); nevertheless, its extraction can still be useful to identify the limiting process. Positive charges in the bottom of C-GaN can either fill a 2D hole gas (2DHG) or occupy defects such as carbon acceptors [15].

-Process 2: *electron current through strain-relief layer*. As the drain-substrate voltage is increased, electrons can be injected from the substrate to the buffer, thus being trapped there. The temperature-dependence of this process depends on the dominant mechanism responsible for vertical leakage. This process leads to negative charge storage.

-Process 3: *hole generation and current through uid layer*. At high voltages, vertical leakage (likely by band-to-band tunneling through the uid layer) can promote the transfer of electrons from the buffer to the 2DEG, with consequent generation of holes in the buffer. Such holes are pushed by the electric field down, till the bottom of the buffer, at the interface with the strain-relief layer, and fill the 2DHG [15]. This process leads to positive charge storage, and its activation energy depends on the dominant vertical leakage processes.

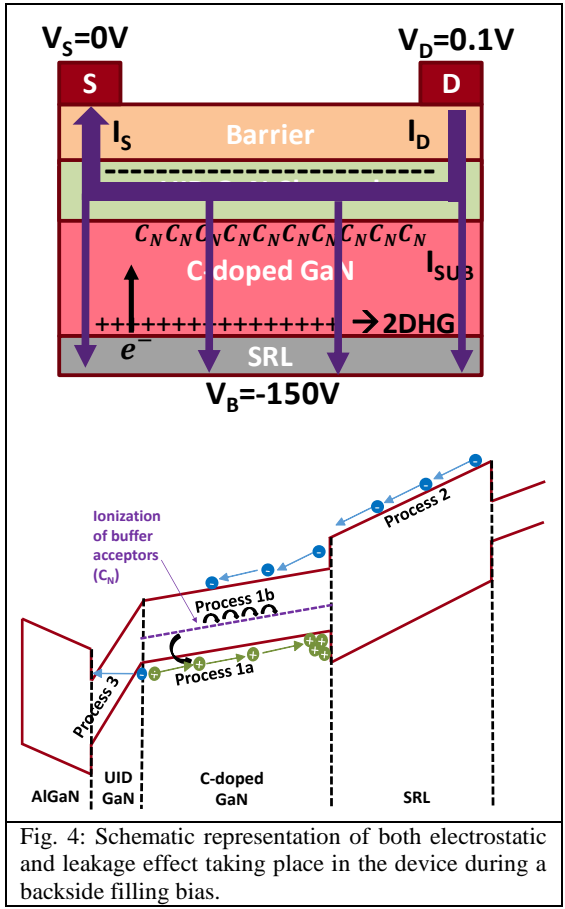

Fig. 4: Schematic representation of both electrostatic and leakage effect taking place in the device during a backside filling bias.

### 3.3 Pulsed backgating current transients

Given the three processes described above, our interest was to understand which one is prominent for each temperature and voltage range. Although with the time-dependent measurements described above we have proven the existence of different processes (responsible for positive and negative charge storage), the identified current transients were too fast for a clear fitting. For this reason, a set of complementary measurement were undertaken using pulsed substrate bias, with a much faster (10 µs) setup. Changes in substrate bias are applied to the silicon, and the substrate bias is periodically interrupted for 10 µs in a logarithmic time sequence in order to measure and record any change in the drain current. Signal sequences as a function of time applied at the drain contact and at the silicon substrate are depicted in Fig.5.

The goal of pulsed back-gating current transient is to detect trapping mechanisms in micro-seconds time frame, avoiding the depletion of channel due to the high electric field. Pulsed $I_D$ transients were recorded in 100s time frame at different substrate voltages, from $V_B$=-10V up to -200V, both in low and high temperature regime (FILLING).

Complementary substrate-bias pulsed experiments are shown in Fig. 6 and 7 (filling/recovery experiments at different voltages, and low/high temperatures).

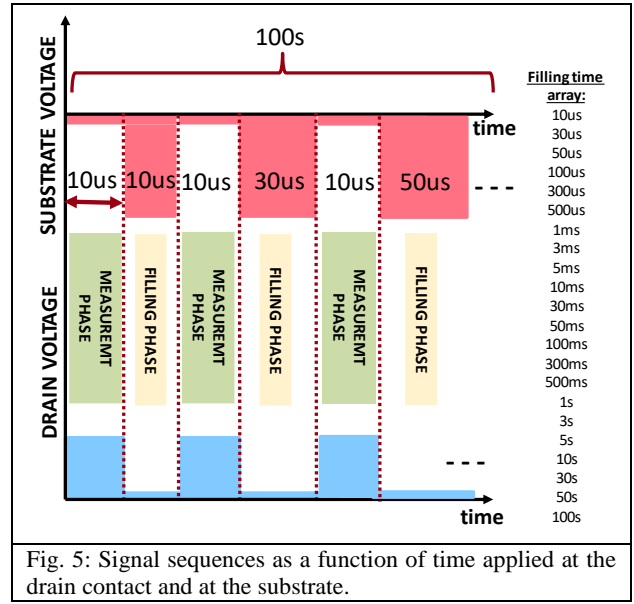

Fig. 5: Signal sequences as a function of time applied at the drain contact and at the substrate.

Fig. 6 shows the drain current variation over time in the low temperature regime, 50 °C. The change in the drain current increases with the voltage applied to the back of the device; current reaches its minimum after $t_{min}$=10s at $V_B$=-200V. For t> $t_{min}$ the device experiences an increase in the current, suppressing the previous negative charge storage. This is consistent with the behavior we have observed experimentally in the previous analysis: two mechanisms are taking place in the structure with two different time constants. The 50 °C recovery measurements indicate that 100 seconds are sufficient to observe full recovery at that temperature.

When comparing measurements collected in the low and high temperature regime (Fig. 7), the data suggests that temperature speeds up both mechanisms. At high temperature, the maximum current drop occurs at $t_{min}$'=0.1s, followed by a return to almost the same current level at $V_B$=0V. This indicates that both mechanisms are fully recoverable and almost no net charge is stored after bias filling. Temperature accelerates the positive charge storage and leads to an almost complete compensation of the overall drain current.

An Arrhenius plot analysis allowed to extract the activation energy for the negative and the positive transients. The negative transient, i.e. decreasing current, can either be ascribed to a) negative charge capture within the structure due to electron leakage through the strain-relief layer (process 2); or to b)

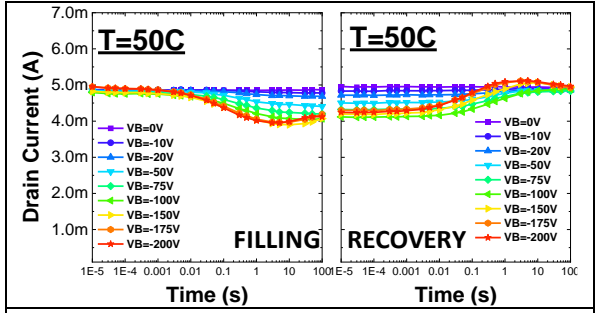

Fig. 6: Pulsed backgating current transients in the low temperature regime.

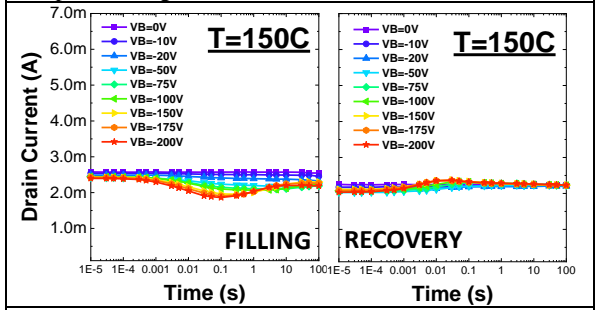

Fig. 7: Pulsed backgating current transients in the high temperature regime.

charge redistribution in C-GaN (process 1).

Although for process 1 the net charge in all layers stays constant, charge redistribution causes negative charges to be closer to the 2DEG than the positive charges, causing a decrease in 2DEG density, i.e. decreasing drain current. Interestingly, the time constant and amplitude of the current decrease fit perfectly to the current increase in the recovery measurements (Fig. 6-7), suggesting a symmetric process. This characteristic fits well to charge redistribution, while electron current through the strain-relief layer could in theory lead to similar stress transients but it cannot explain the observed recovery transient. As the activation energy is roughly 0.3 eV (Fig. 8 (b)), i.e. significantly smaller than the activation energy of carbon acceptors, we consider conduction in defect bands (Process 1b) more likely [14]. Poole-Frenkel lowering of activation energies cannot be ruled out, however at 20-100 V the electric fields are rather small to cause significant $E_A$ lowering. The increase of the current, i.e. the positive transient, has to be related to positive charge accumulation in the buffer. This means that positive charges have to propagate through the uid layer, i.e. process 3. However, interestingly the activation energy decreases significantly with increasing bias from Ea=0.8 eV for VB=-100 V to $E_A$=0.3 eV for VB<-150 V. One could interpret these results as follows: at low biases hole generation limits the transient behavior, and this process has an activation energy of 0.8 eV or even higher. With increasing bias, hole generation accelerates faster than process 1 (hole transport), and for $V_B$<-150 V process 1 is limiting the transient behavior. This explains the saturation of $E_A$ at 0.3 eV, exactly the same value found for process 1 in the decreasing transient. The positive charge accumulation can also be observed in the recovery transients in Fig. 6. For low biases, due to the longer time constants of the hole injection, only charge redistribution takes place during stress. Hence, in the recovery one observes only charge redistribution (in the opposite direction than during stress), i.e. only the increasing transient. For higher biases, on the other hand, holes are generated at the uid layer during stress, as indicated by the current increase. As a result, in the recovery transients after charge redistribution the current is larger than in a fresh device, as in total there is still a net positive charge in the structure. Depleting the structure of these positive charges takes longer, leading to the slower decrease of the current back to the initial value.

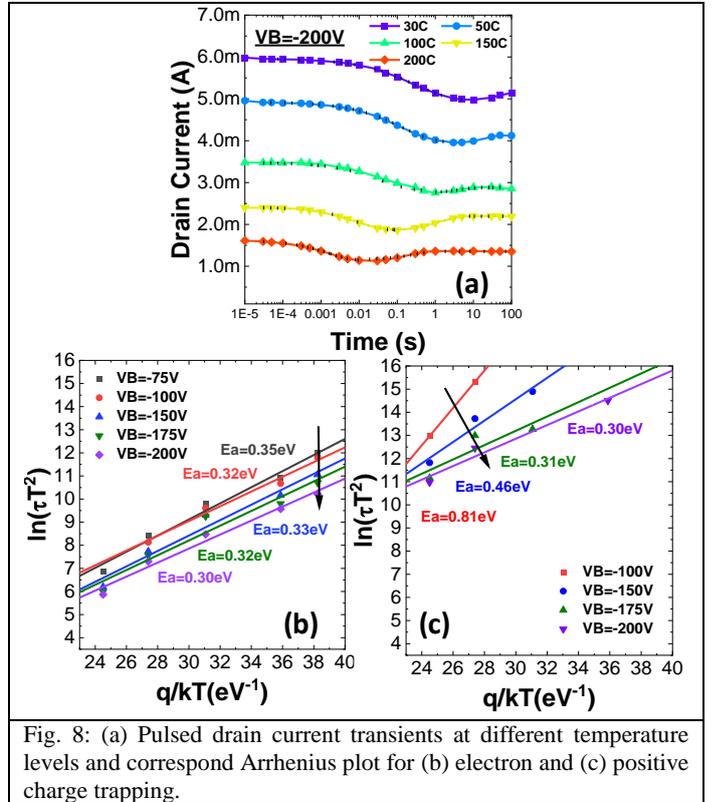

Fig. 8: (a) Pulsed drain current transients at different temperature levels and correspond Arrhenius plot for (b) electron and (c) positive charge trapping.

3. Conclusions

In summary, this paper investigates the kinetics of buffer trapping in GaN-based normally-off high-voltage transistors through a time and temperature dependent analysis. We propose to explain the trapping/de-trapping kinetics by considering the interplay between three processes: 1) the redistribution of charge *in C-doped GaN*. Due to an electric field in the C-doped GaN layer, positive charges will propagate from the top to the bottom of C-doped GaN, with consequent generation of fixed negative charge near the 2DEG and mobile holes on the bottom of the epitaxial stack; 2) the injection of electrons from the substrate to buffer traps, that result in the storage of negative charges in the buffer; 3) the generation of holes and current through the uid layer as well as consequent storage at the bottom of the buffer. We show that the negative charge storage kinetics cannot be described by the 0.8-0.9 eV activation energy, usually ascribed to the ionization of carbon acceptors. A model comprising transport of carriers through defect bands is proposed to explain the measured 0.3 eV activation. In the recovery tests, at low bias only charge redistribution can be observed, resulting in the storage of negative charge near the 2DEG. This leads to an increase in drain current. On the other hand, at high voltages the effect of holes injected through the uid layer results in a net positive charge in the structure, and in non-monotonic recovery kinetics.

The experimental data discussed within this paper indicate that a conventional Arrhenius analysis may fail at describing the trapping/detrapping kinetics in GaN-based HEMTs, since transport phenomena may pose a limit to the trapping and detrapping kinetics.

**This project has received funding from the ECSEL Joint Undertaking (JU) under grant agreement No 826392. The JU receives support from the European Union's Horizon 2020 research and innovation programme and Austria, Belgium, Germany, Italy, Norway, Slovakia, Spain, Sweden, Switzerland**